\documentclass[review,onefignum,onetabnum]{siamart171218}

\usepackage{subcaption}

\usepackage{lipsum}
\usepackage{amsfonts}
 \usepackage{amsmath}
\usepackage{amssymb}

\usepackage{graphicx}
\usepackage{epstopdf}
\usepackage{algorithmic}
\ifpdf
  \DeclareGraphicsExtensions{.eps,.pdf,.png,.jpg}
\else
  \DeclareGraphicsExtensions{.eps}
\fi

{\vskip 2pt \begin{compactitem}[#1]\setlength{\itemsep}{2pt}}
{\end{compactitem}\vskip 2pt}

\newcommand{\be}{\begin{equation}}
\newcommand{\ee}{\end{equation}} 
\newcommand{\lb}{\label}
\newcommand{\OL}{\overline}
\newcommand{\wh}{\widehat}

\newcommand{\const}{({\rm const.})}

\newcommand{\bk}{{\bf k}}

\newcommand{\br}{{\bf r}}
\newcommand{\bu}{{\bf u}}

\newcommand{\bx}{{\bf x}}

\newcommand{\mE}{{\mathcal{E}}}

\newcommand{\bdot}{{\mbox{\boldmath $\cdot$}}}

\newsiamremark{remark}{Remark}
\newsiamremark{hypothesis}{Hypothesis}
\crefname{hypothesis}{Hypothesis}{Hypotheses}
\newsiamthm{claim}{Claim}

\headers{Calculating Spectra by Sequential High-Pass Filtering}{D. Zhao, and H. Aluie}

\title{Calculating Spectra by Sequential High-Pass Filtering 
}

\author{Dongxiao Zhao\thanks{School of Naval Architecture, Ocean and Civil Engineering, Shanghai Jiao Tong University, Shanghai 200240, China and Department of Mechanical Engineering, University of Rochester, Rochester, New York
  (\email{dongxiaozhao@sjtu.edu.cn}).}
\and Hussein Aluie\thanks{Department of Mechanical Engineering, and Department of Mathematics, University of Rochester, Rochester, New York 
  (\email{hussein@rochester.edu}).}}

\usepackage{amsopn}

\makeatletter
\newcommand*{\addFileDependency}[1]{
  \typeout{(#1)}
  \@addtofilelist{#1}
  \IfFileExists{#1}{}{\typeout{No file #1.}}
}
\makeatother

\newcommand*{\myexternaldocument}[1]{%
    \externaldocument{#1}%
    \addFileDependency{#1.tex}%
    \addFileDependency{#1.aux}%
}

\ifpdf
\hypersetup{
  pdftitle={Calculating Spectra by Sequential High-Pass Filtering},
  pdfauthor={D. Zhao, and H. Aluie}
}
\fi

\myexternaldocument{ex_supplement}

\begin{document}
\nolinenumbers
\maketitle

\begin{abstract}
We expand on the method of sequential filtering for
calculating spectra of inhomogeneous fields. Sadek \& Aluie [Phys. Rev. Fluids \textbf{3}, 124610 (2018)] showed that the kernel has to have at least $p$ vanishing moments to extract a power-law spectrum $k^{-\alpha}$ with $\alpha<p+2$ by low-pass filtering.
Here, we show that sequential high-pass filtering allows for extracting steeper spectra with $\alpha<2p+3$ using the same $p$-th order kernel. For example, any spectrum of a field that is shallower than $k^{-5}$ can be extracted by sequential high-pass filtering using any 1st order kernel such as a Gaussian or top-hat. Finally, we demonstrate how second-order structure functions fail to capture spectral peaks because they cannot detect scaling that is too shallow.
\end{abstract}



\section{Introduction}
Charles Meneveau  made foundational contributions to the theory of turbulence and to Large Eddy Simulation (LES) modeling \cite{meneveau1987simple,meneveau1991multifractal,Meneveau94,meneveau1996lagrangian,Cerutti98,bou2005scale,Chevillard08}. He was one of the earliest pioneers who recognized the potential of the coarse-graining (or filtering) framework of Leonard \cite{Leonard75} and Germano \cite{Germano92} to gain insight into the multiscale physics of turbulence \cite{liu1994properties}, including the spatial distribution of energy cascade across scales \cite{meneveau1991dual,meneveau1991analysis}. Here, as in the works of Charles (e.g., \cite{scotti1997fractal,MeneveauKatz00}), we use the terms `filtering' and `coarse-graining' inter-changeably, where the latter emphasizes  the analysis of scale dynamics and has a long history in physics, which goes far beyond mere signal processing that the term `filtering' may suggest.

The filtering framework provides the theoretical basis for subgrid scale (SGS) modeling in LES \cite{MeneveauKatz00}. A primary objective in LES is practical: an accurate SGS model that is numerically stable. Significant advances have been achieved in this regard (e.g., \cite{Piomellietal91,buzzicotti2018energy,deskos2021review,goc2021large,ge2021large}), and the field of LES is arguably mature. The review by Charles Meneveau and Joseph Katz \cite{MeneveauKatz00} on LES has become a classic and remains an invaluable reference approximately 25 years on. Since LES is primarily concerned with SGS modeling, the filtering scale $\ell$ is often taken to be a fixed length of the order of the ‘integral length-scale’ $\ell_0$ in a turbulent flow. This is because the scales $\ell\ll\ell_0$ are expected to be universal in turbulence, justifying general closures for those scales \cite{Frisch95}.

Less common and beyond its LES utility is employing the filtering framework to probe the dynamics at \emph{all} scales (e.g., \cite{Cerutti98,Eyink95,Chenetal03,AluieEyink09}). The idea of using sequential filtering to extract the energy content at different scales; \textit{i.e.},  the spectrum, was recently introduced in \cite{SadekAluie18PRF}. 
Compared to traditional methods, a main advantage of the so-called filtering spectrum is the scale decomposition of a field at any geographic location and any instant of time, without requiring homogeneity. This advantage is shared with the wavelet transform \cite{daubechies1992ten}, which falls within the filtering framework by using a wavelet function as the filtering kernel. In fact, Charles Meneveau was a pioneer in using wavelets to analyze turbulence \cite{meneveau1991dual,meneveau1991analysis}. However, compared to wavelets, simple low-pass filtering can guide subgrid models more naturally and has arguably provided physical insight into the mutiscale dynamics more transparently. This is because simple low-pass filtering partitions the flow into just two sets of scales (larger and smaller than $\ell$), which allows for a tractable analysis of their dynamic coupling as a function of $\ell$ in a manner similar to renormalization-group methods \cite{goldenfeld2018lectures}.

\subsection{Fourier Methods}
By far, the most common method for determining spectra is via the Fourier transform. However, Fourier analysis is fraught with complications when applied to inhomogeneous fields. Afterall, Fourier modes are not an eigenbasis for arbitrary domains and boundary conditions \cite{champeney1987handbook,krantz2019panorama}. Measuring the spectrum via a Fourier transform of the auto-correlation function, sometimes known as the Wiener-Khinchin theorem \cite{champeney1987handbook}, is also not justified in the presence of boundaries or if the field is statistically inhomogeneous such as with a spatially varying mean or autocorrelation. 
In practical applications, Fourier analysis of inhomogeneous fields (or non-stationary temporal signals) is often performed \cite{oppenheim1999signalprocessing,ThomsonEmery01} after removing the ensemble-mean \cite{soulard2012inertial}, detrending \cite{Savage+2017,ORourke2018}, and/or tapering (\textit{i.e.} windowing) \cite{Scott2005,khatri2018surface}. Doing so removes potentially important components of the dynamics. An emblematic example is the global oceanic circulation, for which it had been asserted since the advent of global satellite altimetry in the 1990s that its wavenumber spectrum's peak is at scales $O(100)~$km based on detrended and windowed Fourier analysis (e.g. \cite{FerrariWunsch09,Torresetal2018jgr,Kleinetal2019}). It was recently shown \cite{Storer2022NatComm,buzzicotti2023spatio} that this is untrue and that the spectral peak is in fact at $O(10^4)~$km. The spectral peak and the existence of a power-law scaling over scales $>10^3~$km in the oceanic circulation could not have been detected from windowed Fourier analysis because all scales larger than the window size (typically taken to be a few hundred kilometers to avoid continental boundaries and curvature effects) are implicitly removed. These limitations of Fourier analysis exist for many realistic flows.

\subsection{Structure Functions}
Another tool for analyzing scales is the 2nd-order structure function. It has been a valuable phenomenological tool in turbulence theory, but it requires statistical averaging and is not a formal scale decomposition of a field \cite{Frisch95}. Unlike a spectrum, which when integrated yields total energy (Parseval's relation), a 2nd-order structure function, $S_2(r)$, does not follow such a relation\footnote{While the sum of the 2nd-order structure function and the autocorrelation yields total energy, the sum lacks scale information and is not a scale decomposition. If Fourier analysis is justified, such as for homogeneous flows, it is possible to relate the 2nd-order structure function, $S_2(r)$, to the Fourier spectrum, $E(k)$, but this follows directly from the Wiener-Khinchin relation and involves a weighted average of $E(k)$ over the \emph{entire} $k$-space, $S_2(r)=2\int_0^\infty dk\,(1-cos(k\,r))E(k)$ (e.g. \cite{babiano1985structure,Pope2001}).}. As we shall demonstrate in this paper, at any scale $r$, $S_2(r)$ can have significant contributions from all scales larger or smaller than $r$. This is unsurprising since for a field such as velocity $u(x)$, $S_2(r)=\langle|\delta u(x;r)|^2\rangle$ at scale $r$ is constructed from increments $\delta u(x;r)=u(x+r) - u(x)$ of separation $r$ before spatial averaging, $\langle\dots\rangle$. Increments $\delta u(r)$ can have contributions from all scales larger or smaller than $r$ depending on the regularity (or smoothness) of the field $u(x)$ \cite{Eyink05,Eyink09,Aluie17} (see discussion following eq.~(4) in \cite{aluie2010scale}). It is known that the power-law scaling of a 2nd-order structure function, $S_2(r)\sim r^{\alpha}$, is related to that of the Fourier spectrum, $E(k)\sim k^{-\alpha-1}$, but only if $\alpha<2$, \textit{i.e.} the scaling relation breaks down if $E(k)$ is steeper than $k^{-3}$ as a function of wavenumber $k$ (e.g. \cite{babiano1985structure,biferale2001inverse}). Perhaps less well-known is that the scaling relation between $S_2(r)$ and $E(k)$ also breaks down when $\alpha < 0$, \textit{i.e.} $E(k)$ is shallower than $k^{-1}$ \cite{Eyink95,Eyink05,aluie2010scale}. This prevents the 2nd-order structure function from capturing spectral peaks as we shall demonstrate below. 
Another obvious limitation, shared with Fourier analysis, is that structure functions do not provide spatial information about various scales.

\subsection{Filtering Spectrum}
The so-called filtering spectrum was recently proposed \cite{SadekAluie18PRF} to determine spectral content using straightforward filtering in physical space, which is closely related to the continuous wavelet transform \cite{daubechies1992ten,perrier1995wavelet}. This permits its application to inhomogeneous flows with complex boundaries and allows us to probe scales of both the mean and fluctuating fields concurrently \cite{buzzicotti2023spatio}.
The approach was used to measure the first global energy spectrum of the oceanic general circulation \cite{Storer2022NatComm}. 

The filtering spectrum can be regarded as a generalization of the Fourier spectrum to inhomogeneous fields. The filtering spectrum
is an energy-preserving scale decomposition \cite{SadekAluie18PRF} and can represent the non-quadratic kinetic energy content at different scales of variable-density flows as shown in \cite{Zhao22JFM,zhao2018inviscid,Aluie11c,aluie2012conservative,Aluie13}. It was recently generalized to quantify shape anisotropy at different scales \cite{zhao2023measuring}. Subsequent works \cite{Raietal2021,Zhao22JFM,Storer2022NatComm,buzzicotti2023spatio,juricke2023scale,solodoch2023basin,schubert2023open,loose2023comparing,li2024eddy,liu2024spatial,khatri2024scale,xue2024surface,kouhen2024convective} demonstrate the possibility of performing a meaningful scale decomposition of inhomogeneous fields and determining their spectra, which satisfy both positive semi-definiteness and energy conservation, without the need for the orthogonality structure provided by Fourier modes. If the filtering kernel has a sufficient number of vanishing moments, the filtering spectrum follows any power-law scaling that the Fourier spectrum may have (assuming Fourier analysis is possible). In fact, the filtering spectrum converges to the Fourier spectrum when using a kernel with an infinite number of vanishing moments (e.g. the Dirichlet kernel), which is justified only for homogeneous fields given the highly non-local nature of such kernels in $x$-space.

The filtering spectrum as a method is especially valuable in permitting us to visualize (in physical space) the flow at different scales in a self-consistent manner \cite{buzzicotti2021}. A disadvantage of the filtering spectrum compared to the Fourier spectrum is that it involves smoothing as a function of scale \cite{SadekAluie18PRF}. This is the price paid for gaining spatially local information at different scales and generalizing the notion of a spectrum to non-homogeneous fields. Concurrently exact spatial and scale localization is forbidden by the uncertainty principle \cite{sogge2017fourier,krantz2019panorama}.

\subsection{Paper Outline}
The following section~\ref{sec:Preliminaries} provides preliminaries and a brief review of the low-pass filtering spectrum from \cite{SadekAluie18PRF}. Section~\ref{sec:HPfilteringspectrum} presents the main analytical results, including a method for constructing higher order kernels. 
Section~\ref{sec:numerics} demonstrates the results using numerical data. 
Section~\ref{sec:StructureFunctions} compares filtering spectra and structure functions. The paper concludes with a brief summary and an appendix containing the mathematical derivation of the main result.

\section{Preliminaries} \label{sec:Preliminaries}
In a periodic domain $\bx\in [-L/2,L/2)^n$ in $n-$dimensions, the Fourier transform and its inverse are, respectively
\begin{eqnarray}
\hat{f}(\bk) &=& \frac{1}{L^n}\int_{-L/2}^{L/2} d^n\bx~f(\bx) \,e^{-i\frac{2\pi}{L} \bk\bdot\bx}\\
f(\bx) &=& \sum_\bk  \hat{f}(\bk) \, e^{i\frac{2\pi}{L} \bk\bdot\bx}
\end{eqnarray}
This normalization guarantees that $\hat{f}(\bk=0)$ equals the spatial average, $\langle f(\bx)\rangle = {L^{-n}}\int_{-L/2}^{L/2} d^n\bx~f(\bx) \,$. 
We define the Fourier spectrum of $f(\bx)$ as
\be E(k) = \sum_{k-\frac{1}{2}<|\bk|\le k+\frac{1}{2}} \frac{1}{2}|\hat{f}(\bk)|^2, \hspace{1cm}k = 0,1,2, \dots,
\ee
where $|\bk|=\sqrt{k_x^2+k_y^2+k_z^2}$.
The Fourier coefficients satisfy Plancherel's relation,
\be \langle \frac{1}{2}|f(\bx)|^2\rangle =\sum_{\bk}~\frac{1}{2}|\hat{f}(\bk)|^2 = \sum_{k=0}^{\infty} E(k)~.
\lb{eq:Plancherel}\ee 

\subsection{Filtering}
For any field $u(\bx)$, a coarse-grained or (low-pass) filtered version of this field, which contains spatial variations
at scales $>\ell$, is defined in $n$-dimensional space as \cite{Leonard75,Germano92,MeneveauKatz00,Eyink05}
\be
\OL u_\ell(\bx) = \int \mathrm{d}^n\br~ G_\ell(\bx-\br)\, u(\br).
\lb{eq:filtering_1}\ee
Kernel $G_\ell(\br)= \ell^{-n} G(\br/\ell)$ is the dilated version of the ``parent kernel'' $G(\br)$, which is normalized.  $G_\ell(\br)$ has its main support over a region of diameter $\ell$. 
Operation (\ref{eq:filtering_1}) may be interpreted as a local space average in a region of size $\ell$ centered at point $\bx$. It is, therefore, a scale decomposition performed in x-space that partitions length scales in the system into large ($\gtrsim\ell$), captured by $\OL{u}_\ell$, and small ($\lesssim\ell$), captured by the residual 
\be
u'_\ell=u-\OL{u}_\ell.
\ee
We assume that $\int \br\, G_\ell(\br) \, \mathrm{d}^n \br = 0$, which ensures that local averaging is symmetric. 
\subsection{Low-pass filtering spectrum}
The low-pass filtering spectrum is defined as \cite{SadekAluie18PRF}
\begin{align} 
\OL{E}(k_\ell)\equiv \frac{d}{dk_\ell}\left\langle {\frac{1}{2}|\OL{u}_\ell(\bx)|^2}\right\rangle = -\frac{\ell^2}{L}\frac{d}{d\ell}\left\langle\frac{1}{2} |\OL{u}_\ell(\bx)|^2\right\rangle,
\label{eq:FilteringSpectrum}
\end{align}
where $k_\ell=L/\ell$ is a `filtering wavenumber', $L$ is a characteristic length-scale (e.g. domain size), and $\ell$ is the scale being probed. 
Eq.~\eqref{eq:FilteringSpectrum} measures the energy density (per wavenumber) at scale $\ell$ by varying it and probing the associated variations in coarse energy, $\langle |\OL{\bu}_\ell(\bx)|^2\rangle/2$, which is the cumulative spectrum at \emph{all} scales larger than $\ell$. The main advantage of this method is that it does not rely on Fourier transforms and, therefore, can be easily applied to non-periodic or inhomogeneous data \cite{SadekAluie18PRF,Storer2022NatComm}. In a periodic domain,  Fourier and low-pass filtering spectra agree if $G_\ell$ has sufficient vanishing moments. In fact, the two spectra have an explicit relationship expressed by eq.~(16) in \cite{SadekAluie18PRF}.

\section{High-pass filtering spectrum \lb{sec:HPfilteringspectrum}}
For any given filtering kernel, eq.~(17) of \cite{SadekAluie18PRF} highlights the reason low-pass sequential filtering stops being meaningful if the Fourier spectrum is too steep: large-scale ($k\ll k_\ell$) contributions to $\OL{E}(k_\ell)$ dominate even in the limit of very small filtering scales ($k_\ell\to0$).
Note that the derivation by \cite{SadekAluie18PRF} followed a similar derivation by \cite{perrier1995wavelet} for the wavelet spectrum, which has the same limitation: the lowest order wavelet cannot capture spectral scaling steeper than $k^{-3}$. In this section, will show that high-pass sequential filtering provides significant improvement and allows for capturing steeper spectra using the same kernel by reducing the influence of large scales ($k\ll k_\ell$).

\subsection{Kernel properties}
We shall assume that the kernel is a real-valued even function, $G(\br) = G(-\br)$. Hence, its Fourier transform, $\wh{G}(\bk)$, will also be real-valued. Any spherically symmetric kernel is even. We also assume that the filter kernel is normalized, $\int d\br \,G(\br) = \wh{G}(\bk=0) = 1$.

In practical applications, filtering kernels are often chosen to be sufficiently localized in x-space to avoid prohibitive computational costs. Here, we shall restrict our consideration to kernels that decay faster than any power in x-space,
$G(\br) \le \const r^{-m}$ for any $m$ as $|\br| \to \infty$, where $r = |\br|$. Examples of such kernels include the Gaussian, $(\frac{1}{2\pi})^{n\over2}e^{-|\br|^2/2}$, or kernels that have compact support (i.e. has zero value beyond a finite spatial extent) such as the Top-hat kernel,
\begin{eqnarray}
H_\ell(x)&=&\begin{cases}
    1/\ell, & \text{if $|x|<\ell/2$}.\\
    0, & \text{otherwise}.\\
  \end{cases}\lb{app_eq:Tophat_x}
\end{eqnarray}
These kernels are also useful analytically since the fast decay in x-space guarantees smoothness in k-space. A Taylor-series expansion near the origin in k-space yields:
\be
\wh{G}(k) = \wh{G}(0) + k\,\wh{G}^{(1)}(0) + k^2\,\frac{\wh{G}^{(2)}(0)}{2!} + \dots
\lb{eq:TaylorExpandKernel_1}\ee
where $f^{(n)}(s)$ denotes the $n$-th derivative, $\frac{\partial^n}{\partial s^n} f(s)$.

Moments of a kernel are related to its derivatives in k-space:
\be \int^{+\infty}_{-\infty} dx ~x^n \,G(x) = \wh{G}^{(n)}(k)\Big|_{k=0}.
\lb{eq:MomentsDerivatives}\ee
Since even kernels, $G(x)=G(-x)$, have vanishing odd moments, it follows from eq.~(\ref{eq:MomentsDerivatives}) that 
 $ \wh{G}^{(n)}(0)=0$ for all odd integers $n$ for any even kernel $G(x)$. 

We shall call a kernel $G(x)$ ``$p$-th order'' iff
\begin{eqnarray} 
 \int^{+\infty}_{-\infty}  dx ~x^n G(x) &=& 0 \hspace{1cm} \mbox{for} \hspace{1cm} n=1,\dots,p, \nonumber\\
 \mbox{and}\hspace{1cm} \int^{+\infty}_{-\infty}  dx ~x^{p+1} G(x) &\ne& 0.
\lb{eq:pthOrderKernel}
\end{eqnarray} 
Any even kernel is of an odd integer order $p\ge 1$. For example, the Gaussian and Top-hat kernels are of order $p=1$. As discussed in \cite{SadekAluie18PRF}, the order of the kernel is a key property for extracting the correct spectrum by sequential filtering. 

For a normalized even $p$-th order kernel, the Taylor expansion in eq. (\ref{eq:TaylorExpandKernel_1}) becomes
\begin{eqnarray} 
\wh{G}(k) &=& 1 + k^{p+1}\underbrace{\left[\frac{\wh{G}^{(p+1)}(0)}{(p+1)!}  + k^{2}\,\frac{\wh{G}^{(p+3)}(0)}{(p+3)!}  + \dots \right]}_{\phi(k)}, \lb{eq:TaylorExpandKernel}
\end{eqnarray} 
where
\begin{eqnarray} 
\phi(0) = \const\ne 0. 
\lb{eq:PhiProperty0}\end{eqnarray}

Note that in the Taylor expansion in eq. (\ref{eq:TaylorExpandKernel}), we are using smoothness properties (in k-space) of the kernel and not of the field being filtered. 

\subsection{Constructing high-order kernels from a Gaussian\lb{sec:Gp3kernel}}
\begin{figure}[htbp]
  \centering
  \includegraphics{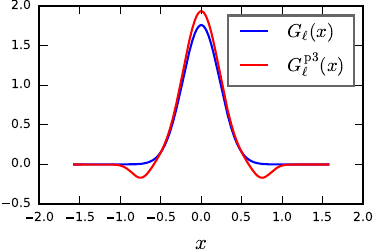}
  \caption{Gaussian kernel $G_\ell(x)$ in eq.~\eqref{eq:Gaussian} and the 3rd-order kernel $G^\mathrm{p3}_\ell(x)$ in eq.~\eqref{eq:GaussP3} with parameters $c=1.1, \ell/\ell'=2$. The domain is $[-\pi, \pi]$ and the filtering scale is $\ell=\pi/4$.}
  \label{fig:kernel_1D}
\end{figure}

The Gaussian kernel is a 1st order kernel ($p=1$). For the numerical results below, we use the Gaussian kernel in $n$-dimensions
\be
    G_\ell(\br) = \left(\frac{6}{\pi\ell^2}\right)^{\frac{n}{2}}~e^{-6|\br|^2/\ell^2}
\lb{eq:Gaussian}\ee
Motivated by the procedure in \cite{SadekAluie18PRF} to build higher order kernels, we can construct $p$th-order kernels by a linear combination of $(p+1)/2$ Gaussians with different filtering widths and different center locations. For example, to construct a $p=3$ kernel of width $\ell$ we use two Gaussians of width $\ell$ and $\ell'$, respectively,
\be
    G^\mathrm{p3}_\ell(x) \equiv c\ G_\ell(x) - c'\  G_{\ell'}(x-x_0) - c'\ G_{\ell'}(x+x_0)~.
\lb{eq:GaussP3}\ee
Here, $c, c', x_0$ are dilation and translation parameters to be determined from the properties of $G^\mathrm{p3}_\ell$:
\be
    \begin{cases}
        \int_{-\infty}^\infty G^\mathrm{p3}_\ell(x) dx &= 1, \\
        \int_{-\infty}^\infty x^2\ G^\mathrm{p3}_\ell(x) dx &= 0.
    \end{cases}
\ee
The four free parameters $\ell', c, c', x_0$ are thus reduced to two, 
\be
    c' = \frac{c-1}{2},\quad x_0 = \frac{c\ \ell^2}{12(c-1)}-\frac{\ell'^2}{12}
\ee
Figure \ref{fig:kernel_1D} shows $G_\ell(x)$ in eq.~\eqref{eq:Gaussian} and $G^\mathrm{p3}_\ell(x)$ in eq.~\eqref{eq:GaussP3} with the parameters $c=1.1, \ell/\ell'=2$.

\subsection{High-pass Sequential Filtering}
For a field $u(\bx)$, we define the high-pass filtering spectrum as,
\begin{equation} \label{eq:filtering_spec_subscale}
    \OL{E}'(k_\ell)\equiv -\frac{d}{d k_\ell}\langle |u'_\ell(\bx)|^2 \rangle /2 = \frac{\ell^2}{L}\frac{d}{d\ell}\langle |u'_\ell(\bx)|^2 \rangle /2~.
\end{equation}
We now follow the analysis in \cite{SadekAluie18PRF} to characterize the scaling of $\OL{E}'(k_\ell)$. This is pertinent to determine $\OL{E}'(k_\ell)$ is meaningful in the sense that it captures the scaling of the Fourier spectrum when the latter is possible to calculate. Assume that $E(k)\propto k^{-\alpha}$ over $k_a<k<\infty$ for an arbitrary wavenumber $k_a$, then
\begin{equation}
    \begin{split}
        \OL{E}'(k_\ell) &= -\int_0^\infty dk \left|1-\widehat{G}\left(\frac{k}{k_\ell}\right)\right|^2 E(k)\\
        &= \underbrace{-\int_0^{k_a} dk\frac{d}{dk_\ell}\left|1-\widehat{G}\left(\frac{k}{k_\ell}\right)\right|^2 E(k)}_{\mathrm{term~I}\sim k_\ell^{-2p-3}} \underbrace{-\int_{k_a}^\infty dk\frac{d}{dk_\ell}\left|1-\widehat{G}\left(\frac{k}{k_\ell}\right)\right|^2 E(k)}_{\mathrm{term~II}\sim k_\ell^{-\alpha}}~.
    \end{split}
\lb{eq:scalingSpectrum}
\end{equation}
The derivation is in the Appendix. Eq.~\eqref{eq:scalingSpectrum} implies that if the Fourier spectrum has a power-law scaling $E(k)\sim k^{-\alpha}$ at high wavenumbers, then the high-pass filtering spectrum obtained by filtering with a $p$-th order kernel scales as 
\begin{eqnarray}
\OL{E}'(k)&\sim&\begin{cases}
    k^{-\alpha}, & \text{if $\alpha < 2p+3$}\\
    k^{-(2p+3)}, & \text{if $\alpha > 2p+3$}\\
  \end{cases}\lb{eq:FilterSpectrumScaling}
\end{eqnarray}
Therefore, if the Fourier spectrum decays faster than $k^{-(2p+3)}$, the small wavenumber contributions in “term I” of eq.~\eqref{eq:scalingSpectrum} dominate at large $k_\ell$, whereas if $\alpha<2p+3$, then the high-pass filtering spectrum is meaningful in the sense that it can capture the power-law scaling of the Fourier spectrum. This is a significant improvement over the low-pass filtering spectrum presented in \cite{SadekAluie18PRF}, which can only capture power-laws with $\alpha<p+2$.

The steeper is the underlying spectrum, the higher is the order of the filtering kernel required for extracting such a spectrum.
For example, the Gaussian or Top-hat functions are 1st-order kernels and can only extract power-law spectra shallower than $k^{-5}$. A practical consequence of eq. (\ref{eq:FilterSpectrumScaling}) is that if a filtering spectrum is measured using a $p$-th order kernel and exhibits a scaling shallower than $k^{-(2p+3)}$, then the user can have confidence that it reflects the scaling of the Fourier spectrum correctly. Otherwise, if it scales $\sim k^{-(2p+3)}$, then a higher order filtering kernel is required.

The high-pass filtering spectrum is an energy-preserving scale decomposition. Indeed, it is straightforward to verify that its integral yields the total energy:
\begin{eqnarray}
\frac{1}{2}\left\langle |u|^2 \right\rangle = \frac{1}{2}|\left\langle {u} \right\rangle|^2+\int_{0}^\infty dk_\ell~\OL{E}'(k_\ell)~.
\lb{eq:EnergyConservation_1}\end{eqnarray}
In practice, since the spectrum itself is calculated from the \emph{cumulative} high-pass spectrum, $\mE'(k_\ell) \equiv \langle |u'_\ell(\bx)|^2 \rangle /2$, the total energy is retrieved by taking the limit of large filter scale,
\be \lim_{k_\ell\to 0}\mE'(k_\ell) = \frac{1}{2}\left\langle |u|^2 \right\rangle~.
\lb{eq:EnergyConservation_2}\ee

\section{Numerical results}
\begin{figure}[htbp]
  \centering
  \includegraphics{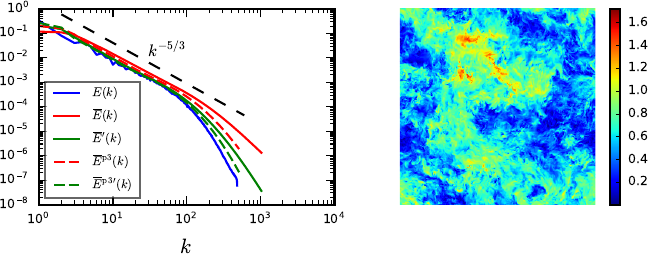}
  \caption{Right panel is a visualization of velocity magnitude from a 2D slice of the 3D isotropic turbulence data from the JHU Turbulence database. The domain is periodic. Left panel shows the Fourier spectrum $E(k)$ alongside the low-pass ($\OL{E}(k)$ and $\OL{E}^{\mathrm{p3}}(k)$) and high-pass ($\OL{E}^{'}(k)$ and $\OL{E}^{'\mathrm{p3}}(k)$) filtering spectra using $G_\ell$ and $G^\mathrm{p3}_\ell(x)$ shown in Fig.~\ref{fig:kernel_1D}. Since the Fourier spectrum over the inertial range has $E(k)\sim k^{-5/3}$, it is sufficiently shallow to be captured by all filtering spectra, including when using a 1st-order kernel. Differences between the spectra can be seen in the dissipation range where the Fourier spectrum is steep.
  }
  \label{fig:JHU_spec}
\end{figure}

\lb{sec:numerics}
Here, we compare the scaling of $\OL{E}'(k_\ell)$ using 1st-order and 3rd-order kernels to the traditional Fourier spectrum.
We use the 3D isotropic turbulence dataset from the Johns Hopkins database \cite{JHU08}, which exists primarily thanks to Charles Meneveau's bold vision and tireless work. Figure~\ref{fig:JHU_spec} shows the Fourier spectrum of a 2D slice, which is periodic. Figure~\ref{fig:JHU_spec} also shows low-pass and high-pass filtering spectra using $G_\ell$ and $G^\mathrm{p3}_\ell(x)$ in Fig.~\ref{fig:kernel_1D}. Since the Fourier spectrum over the inertial range
in Fig.~\ref{fig:JHU_spec} follows $E(k)\sim k^{-5/3}$, it is sufficiently shallow to be captured by all filtering spectra, including when using a 1st-order kernel. Differences appear in the dissipation range, when $E(k)$ becomes too steep. Over those small scales, we see that the high-pass filtering spectrum, $\OL{E}^{'\mathrm{p3}}(k)$, using the 3rd-order kernel $G^\mathrm{p3}_\ell(x)$ is the most accurate. This is in accord with our analytical result above, which indicates that $\OL{E}^{'\mathrm{p3}}(k)$ should capture scaling shallower than $k^{-9}$. When using a 1st-order kernel, we see in Fig.~\ref{fig:JHU_spec} that $\OL{E}^{'}(k)$ is less accurate since it can only capture scaling shallower than $k^{-5}$. Its scaling is comparable to that of the low-pass filtering spectrum $\OL{E}^{\mathrm{p3}}(k)$ using the 3rd-order kernel, which can also capture scaling shallower than $k^{-5}$. We suspect that the slight improvement of $\OL{E}^{'}(k)$ over $\OL{E}^{\mathrm{p3}}(k)$ in Fig.~\ref{fig:JHU_spec} may be due to the reduced influence of large-scales when  
high-pass filtering. Finally, the low-pass filtering spectrum, $\OL{E}(k_\ell)$, using the 1st-order kernel is the least accurate since it can only capture scaling shallower than $k^{-3}$.

\begin{figure}
\centering
\begin{minipage}[b]{1.0\textwidth}
\centering
    \begin{subfigure}{0.4\textwidth}   
    \centering
    \includegraphics[width=0.98\textwidth]{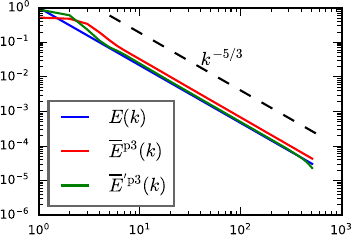} 
        \caption{}
    \end{subfigure} 
    \begin{subfigure}{0.4\textwidth}   
    \centering
    \includegraphics[width=0.98\textwidth]{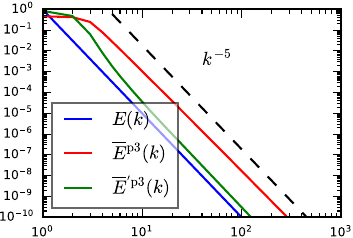} 
        \caption{}
    \end{subfigure}   \\
    \begin{subfigure}{0.4\textwidth}   
    \centering
    \includegraphics[width=0.98\textwidth]{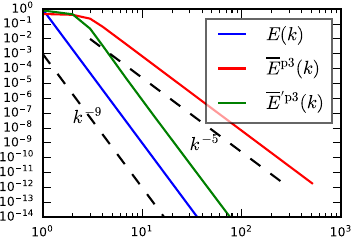} 
        \caption{}
    \end{subfigure} 
    \begin{subfigure}{0.4\textwidth}   
    \centering
    \includegraphics[width=0.98\textwidth]{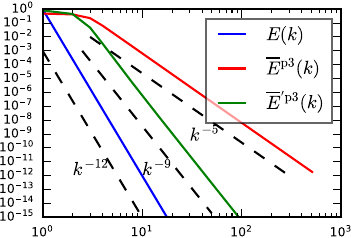} 
        \caption{}
    \end{subfigure} 
\end{minipage}
    \caption{To demonstrate the limitations of low-pass (red) and high-pass (green) filtering spectra, we use 2D periodic fields $\phi(\bx)$ having Fourier spectra $E(k)$ (blue) with scaling exponents $-5/3, -5, -9, -12$ shown panels~(a)-(d), respectively. (a) Both $\OL{E}^\mathrm{p3}(k)$ and $\OL{E}^{'\mathrm{p3}}(k)$ are accurate at capturing $E(k)\sim k^{-5/3}$.
    (b) $\OL{E}^\mathrm{p3}(k)$ locks in at $k^{-5}$ while $\OL{E}^{'\mathrm{p3}}(k)$ remains accurate at capturing $E(k)\sim k^{-5}$.
    (c) $\OL{E}^\mathrm{p3}(k)$ is locked at $k^{-5}$ while $\OL{E}^{'\mathrm{p3}}(k)$ locks in at $k^{-9}$ and stops being accurate at capturing $E(k)\sim k^{-9}$.
    (d) Since $\OL{E}^\mathrm{p3}(k)$ locks at $k^{-5}$ and $\OL{E}^{'\mathrm{p3}}(k)$ locks at $k^{-9}$, neither can capture $E(k)\sim k^{-12}$.
 \label{fig:syntheticdata}}
\end{figure}

We also use 2D synthetic data with Fourier spectra having prescribed power-laws. This is to demonstrate how filtering spectra cannot capture the true spectral scaling if it is too steep for the kernel used, instead locking at the power-law scaling derived in eq.~\eqref{eq:scalingSpectrum}.
We generate several 2D periodic fields $\phi(\bx)$ having Fourier spectra with scaling exponents $-5/3, -5, -9, -12$ shown in Fig.~\ref{fig:syntheticdata}(a)-(d), respectively. As discussed above, the low-pass filtering spectrum $\OL{E}^\mathrm{p3}(k)$ with a 3rd-order kernel can capture a Fourier spectrum shallower than $k^{-5}$, while the high-pass filtering spectrum $\OL{E}^{'\mathrm{p3}}(k)$ can capture spectra shallower than $k^{-9}$. Figure~\ref{fig:syntheticdata} demonstrates our results numerically.

\section{Structure Functions}
\label{sec:StructureFunctions}
As discussed in the introduction, the 2nd-order structure function $S_2(r)$ is a common tool in turbulence \cite{Frisch95} that is often used as a proxy for the Fourier spectrum \cite{Alexakis18,zhou2021turbulence}. It played a central role in Kolmogorov's formulation of his theory \cite{Kolmogorov41,Kolmogorov41b}. It is usually defined as 
\be
S_2(\br) = \langle|\delta u(\bx;\br)|^2\rangle,
\lb{eq:secondstructure}
\ee
where
\be
\delta u(\bx;\br) = u (\mathbf{x}+\mathbf{r}) - u (\mathbf{x}),
\lb{eq:increment}
\ee
is an increment of separation $\br$. Without boldface $\br$, it should be understood that $S_2(r)$ is obtained from $S_2(\br)$ after averaging over all angles.

\begin{figure}[htbp]
  \centering
  \includegraphics[width=0.32\textwidth]{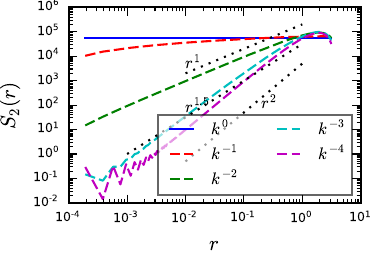}
  \includegraphics[width=0.32\textwidth]{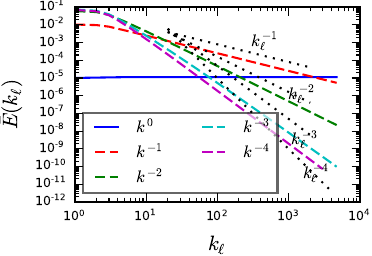}
  \includegraphics[width=0.32\textwidth]{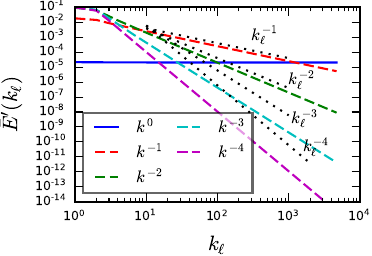}
 \caption{{\bf Left panel:}  power-law scaling of a 2nd-order structure function, $S_2(r)\sim r^{\alpha-1}$, is related to scaling of the Fourier spectrum $E(k)\sim k^{-\alpha}$ (legend), but only when $1<\alpha<3$. We see that the scaling of $S_2(r)$ is no longer related to that of the Fourier spectrum when $E(k)$ is steeper than $k^{-3}$ (magenta) or shallower than $k^{-1}$ (blue). The fields analyzed here are 1D periodic data similar to those shown in Fig.~24 of \cite{zhao2023measuring}.
    {\bf Middle panel:} Low-pass filtering spectra $\OL{E}(k_\ell)$ using a Gaussian kernel applied to the same data used in the left panel. We see that the filtering spectrum has the same scaling as the Fourier spectrum, $\OL{E}(k)\sim E(k) \sim k^{-\alpha}$, for $\alpha<3$. Specifically, it can correctly capture power-law scaling that is shallower than $k^{-1}$ (blue) but fails for power-law scaling steeper than $k^{-3}$ (magenta) since the Gaussian kernel ($p=1$) we are using to calculate $\OL{E}(k_\ell)$ is a first-order kernel. It is possible for $\OL{E}(k_\ell)$ to correctly capture  power-laws steeper than $k^{-3}$ by using a higher-order kernel  \cite{SadekAluie18PRF}. {\bf Right panel:} similar to middle panel, but shows high-pass filtering spectra, which can also capture steeper spectra (magenta) accurately.
 }
  \label{fig:StructureFunctionsCompare}
\end{figure}

While $S_2(r)$ has been a valuable phenomenological tool in turbulence theory, unfortunately it is not a formal scale decomposition of a field \cite{Frisch95}. It is known that its power-law scaling, $S_2(r)\sim r^{\alpha-1}$, is related to that of the Fourier spectrum, $E(k)\sim k^{-\alpha}$, but only if $1<\alpha<3$ \cite{babiano1985structure,biferale2001inverse,Eyink05,aluie2010scale}. This fact is demonstrated in Fig.~\ref{fig:StructureFunctionsCompare} (left panel), where we can see that the $S_2(r)$ power-law scaling can only be between $r^{0}$ and $r^{2}$, corresponding to a Fourier spectral exponent $1<\alpha<3$. Therefore, $S_2(r)$ cannot capture spectral scaling that is either too steep or too shallow.
In contrast, neither low-pass nor high-pass filtering spectra has limitations for capturing spectral scaling that is too shallow, even when using a 1st-order kernel as shown in Fig.~\ref{fig:StructureFunctionsCompare} (middle and right panels).

\begin{figure}[htbp]
  \centering
   \includegraphics{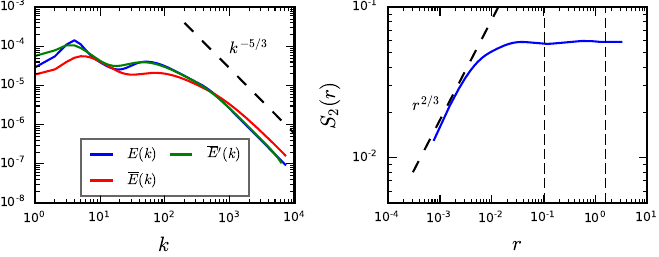}
 \caption{  {\bf Left panel:} Fourier spectrum (blue), along with low-pass (red) and high-pass (green) filtering spectra extracted from the same 1D field using a Gaussian kernel in a periodic domain of size $L=2\pi$. The field has two spectral peaks, which the filtering spectra can capture.
 {\bf Right panel:}  Second-order structure function, $S_2(r)$, calculated from the same field is able to capture the power-law scaling at small scales $r$ but fails to detect the two spectral peaks. This is because $S_2(r)$ cannot capture the shallow spectral scaling that occurs around scales of the spectral peak and larger (\textit{i.e.} smaller $k$) as demonstrated in Fig.~\ref{fig:StructureFunctionsCompare}. Vertical lines indicate scales $r = L / k$ at wavenumbers $k$ where the Fourier spectrum peaks.
 }
  \label{fig:StructureFunctionsPeaks}
\end{figure}

The limitation of $S_2(r)$ in capturing spectra shallower than $k^{-1}$ implies that $S_2(r)$ fails to detect spectral peaks. This is because the spectral slope at the peak follows $k^{0}$. Moreover, $S_2(r)$ cannot capture the scaling at wavenumbers $k$ smaller than that of the peak where the spectral slope is positive. These considerations are demonstrated numerically in Fig.~\ref{fig:StructureFunctionsPeaks}, where we can see that $S_2(r)$ follows the expected $r^{2/3}$ power-law scaling (corresponding to $E(k)\sim k^{-5/3}$) at small scales, but saturates at large scales without capturing the two peaks present in the Fourier spectra. A similar double peaked spectrum was measured in the global ocean circulation \cite{Storer2022NatComm}, where it is believed to be due to different forcing mechanisms at different scales \cite{storer2023global}. Fig.~\ref{fig:StructureFunctionsPeaks} shows that both low-pass and high-pass filtering spectra can capture the Fourier spectrum reasonably well using a 1st-order kernel because they are not limited by shallow scaling.

\section{Conclusions}
\label{sec:conclusions}
We have shown that the spectrum of a field can be extracted by sequential high-pass filtering in physical space. The approach brings significant improvements over sequential low-pass filtering introduced in earlier work \cite{SadekAluie18PRF} because it can capture much steeper spectra. Even when using the lowest order filtering kernel, sequential high-pass filtering can capture spectra shallower than $k^{-5}$. In comparison, both sequential low-pass filtering and wavelet transform can only capture spectra shallower than $k^{-3}$ \cite{SadekAluie18PRF,perrier1995wavelet}. The improvement is rooted in the enhanced insulation from the largest (and most energetic) scales when high-pass filtering.

We also demonstrated how second-order structure functions fail to capture spectral peaks because they cannot detect scaling that is too shallow. This limitation is not shared by either low-pass or high-pass filtering spectra. We note that a high-pass filtered field and its increments are related by 
\be
u'_\ell(\bx) =  u(\bx) - \OL{u}_\ell(\bx) = -\int \mathrm{d}^{n}\br\, G_\ell(-\br) \, \delta {u}(\bx;\br)~. 
\lb{eq:highpassincrements} \ee
This relation implies that a high-pass filtered field $u'_\ell(\bx)$ is exactly equal to the spatial average of increments $\delta {u}(\bx;\br)$ originating from $\bx$ with separations $\br$ within a ball of radius $\sim\ell/2$. That the high-pass filtering spectrum, which is a spatial average of $|u'_\ell(\bx)|^2$ in eq.~\eqref{eq:filtering_spec_subscale} is superior to $S_2(r)$, which is a spatial average of $|\delta {u}(\bx;\br)|^2$ in eq.~\eqref{eq:secondstructure}, underscores the importance of local averaging of increments in eq.~\eqref{eq:highpassincrements}.

\clearpage
\appendix
\section{High-pass Filtering Spectrum Scaling} 
Here, we derive the scaling of $\OL{E}'(k_\ell)$ in eq.~(\ref{eq:FilterSpectrumScaling}). Assume that $E(k)\propto k^{-\alpha}$ over $k_a<k<\infty$, then
\begin{equation}
    \begin{split}
        \OL{E}'(k_\ell) &= -\int_0^\infty dk \left|1-\widehat{G}\left(\frac{k}{k_\ell}\right)\right|^2 E(k)\\
        &= \underbrace{-\int_0^{k_a} dk\frac{d}{dk_\ell}\left|1-\widehat{G}\left(\frac{k}{k_\ell}\right)\right|^2 E(k)}_{\mathrm{I}} \underbrace{-\int_{k_a}^\infty dk\frac{d}{dk_\ell}\left|1-\widehat{G}\left(\frac{k}{k_\ell}\right)\right|^2 E(k)}_{\mathrm{II}}
    \end{split}
\end{equation}
which is split into two terms with contribution from low (term I) and high (term II) wavenumber components. Term II can be recast as
\begin{displaymath}
        \mathrm{II} = k_\ell^{-\alpha} \int_{k_a/k_\ell}^\infty ds \frac{d}{ds}|1-\widehat{G}(s)|^2 s^{1-\alpha}
\end{displaymath}
with $s=k/k_\ell$. Under mild smoothness and decay conditions on $\wh{G}\left(s\right)$, the integral on the right hand side converges to a constant and term II scales as $k_\ell^{-\alpha}$.

For term I, if the filtering kernel is an even $p$th-order kernel, we can recast $\widehat{G}(k)$ using Taylor series expansion \cite{SadekAluie18PRF} 
\begin{displaymath}
    \widehat{G}(k) = 1+k^{p+1}\phi(k)~,\qquad |1-\widehat{G}(\frac{k}{k_\ell})|^2 = (\frac{k}{k_\ell})^{2p+2}\phi^2(\frac{k}{k_\ell})~,
\end{displaymath}
where $\phi(k)$ was defined in eq.~\eqref{eq:TaylorExpandKernel}. Thus 
\begin{equation} \label{eq:term_I_scaling}
\begin{split}
        \mathrm{I} &= -\int_0^{k_a} dk \frac{d}{dk_\ell} \left[\left(\frac{k}{k_\ell}\right)^{2p+2}\phi^2\left(\frac{k}{k_\ell}\right)\right]E(k)\\
        &=-\int_0^{k_a} dk\ k^{2p+2}\left[-(2p+2)\frac{1}{k_\ell^{2p+3}}\phi^2\left(\frac{k}{k_\ell}\right)-\frac{1}{k_\ell^{2p+2}}2\phi\left(\frac{k}{k_\ell}\right)\phi'\left(\frac{k}{k_\ell}\right)\frac{k}{k_\ell^2}\right]E(k)\\
        &=\underbrace{\int_0^{k_a}dk\ (2p+2) k^{2p+2}\phi^2\left(\frac{k}{k_\ell}\right) \frac{1}{k_\ell^{2p+3}} E(k)}_{\mathrm{I_a}} + \underbrace{\int_0^{k_a}dk\ \frac{2\  k^{2p+3} \phi\left(\frac{k}{k_\ell}\right) \phi'\left(\frac{k}{k_\ell}\right) }{k_\ell^{2p+4}}E(k)}_{\mathrm{I_b}}\\
        &\sim k_\ell^{-2p-3}
\end{split}
\end{equation}
The last line holds since $\mathrm{I_a}=k_\ell^{-2p-3} \mathrm{(const.)} \int_0^{k_a} (2p+2)k^{2p+2} E(k) dk\sim k_\ell^{-2p-3}$ and $\mathrm{I_b}$ becomes negligible for $k/k_\ell\to 0$ due to $\phi(0) = \mathrm{(const.)}$ and $\phi'(0) = 0$.

Finally, we have
\begin{equation} \label{eq:subscale_spec_scaling}
    \OL{E}'(k_\ell) = \underbrace{-\int_0^{k_a} dk\frac{d}{dk_\ell}|1-\widehat{G}(\frac{k}{k_\ell})|^2 E(k)}_{\sim k_\ell^{-2p-3}} \underbrace{-\int_{k_a}^\infty dk\frac{d}{dk_\ell}|1-\widehat{G}(\frac{k}{k_\ell})|^2 E(k)|}_{\sim k_\ell^{-\alpha}}
\end{equation}

\section*{Acknowledgments}
This research was funded by US NSF grants PHY-2206380 and OCE-2123496. 
DZ was also supported by the National Natural Science Foundation of China (No.~12202270), and the China Postdoctoral Science Foundation (No.~2023M742269).
HA was also supported by US DOE grants DE-SC0020229, DE-SC0014318, DE-SC0019329, US NSF grant PHY-2020249.
Computing time was provided by NERSC under Contract No. DE-AC02-05CH11231, and the Texas Advanced Computing Center (TACC) at The University of Texas at Austin, under ACCESS allocation grant EES220052.

\bibliographystyle{ieeetr}
\bibliography{references}
\end{document}